\documentclass[pra,onecolumn,notitlepage,amsmath,nofootinbib,superscriptaddress,showkeys]{revtex4-1}

\usepackage[T1]{fontenc}
\usepackage[latin9]{inputenc}
\usepackage{amsmath}
\usepackage{amssymb}
\usepackage{esint}
\usepackage[brazil,english]{babel}
\usepackage{graphicx}
\usepackage[usenames,dvipsnames]{color}
\usepackage{longtable}
\usepackage{dcolumn}
\usepackage{ulem}
\usepackage{nicefrac}

\begin{document}

\title{Classical-quantum correspondence for two-level pseudo-Hermitian systems}

\author{K. Raimundo}
\email{kesley@uel.br}
\author{M. C. Baldiotti}
\email{baldiotti@uel.br}
\affiliation{Departamento de F\'{\i}sica, Universidade Estadual de Londrina, 86051-990, Londrina-PR, Brazil.}

\author{R. Fresneda}
\email{rodrigo.fresneda@ufabc.edu.br}
\affiliation{Universidade Federal
do ABC, Av. dos Estados 5001, 09210-580, Santo Andr\'{e}-SP, Brazil.}

\author{C. Molina}
\email{cmolina@usp.br}
\affiliation{Escola de Artes, Ci\^{e}ncias e Humanidades, Universidade de S\~{a}o Paulo, Av. Arlindo Bettio 1000, CEP 03828-000, S\~{a}o Paulo-SP, Brazil.}

\begin{abstract}
In this work, a classical-quantum correspondence for two-level pseudo-Hermitian systems  is proposed and analyzed. We show that the presence
of a complex external field can be described by a pseudo-Hermitian Hamiltonian
if there is a suitable canonical transformation that links it to a real field.
We construct a covariant quantization scheme which maps canonically related
pseudoclassical theories to unitarily equivalent quantum realizations, such
that there is a unique metric-inducing isometry between the distinct Hilbert
spaces. In this setting, the pseudo-Hermiticity condition for the operators
induces an involution which guarantees the reality of the corresponding
symbols, even for the complex field case. We assign a physical meaning for the
dynamics in the presence of a complex field by constructing a classical
correspondence. As an application of our theoretical framework, we propose a
damped version of the Rabi problem and determine the configuration of the
parameters of the setup for which damping is completely suppressed. The experimental viability of the proposal is studied within a specific context. We suggest that the main theoretical results developed in the present work could be experimentally verified.

\end{abstract}

\maketitle

\section{Introduction}

The simplest system with nontrivial dynamics that we can build in quantum
mechanics is the two-level system. But despite its simplicity, one cannot
underestimate the power of this setup. For instance, two-level models are the
best understood quantum systems and adequately describe several physically
relevant scenarios. Moreover, they play an important role in the understanding
of more intricate arrangements. In general, one can treat a quantum two-level
system as a spin-$\nicefrac{1}{2}$ particle interacting with an external
magnetic field if the spatial dynamics is not taken into account. Thus, a
two-level system is governed by the Pauli equation in $(0+1)$ dimension,
\begin{equation}
\mathrm{i}\frac{\partial v}{\partial t}=\hat{H}v\text{, with }\hat{H} 
= \frac{\boldsymbol{\sigma}}{2}\cdot\mathbf{F}\ \text{and}\ v=\left(
\begin{array}[c]{c}%
v_{1}(t)\\
v_{2}(t)
\end{array}
\right)  \, . \label{SE}%
\end{equation}
In Eq.~(\ref{SE}), $v$ is a two-component spinor, $\boldsymbol{\sigma}%
=(\sigma_{1},\sigma_{2},\sigma_{3})$ are the Pauli matrices and $\mathbf{F}%
=(F_{1}(t),F_{2}(t),F_{3}(t))$ represents an external
field.%
\footnote{\label{f1}We are setting $\gamma=-1$, where $\gamma=gq/2m$
with $q$, $m$ and $g$ being, respectively, the charge, mass and the $g$-factor
of the spin-$\nicefrac{1}{2}$ particle. Also, in this description,
$\mathbf{F}$ has dimension of energy.}%
Therefore, solving a two-level system
is equivalent to solving~(\ref{SE}), to which will be referred as the
\textit{spin equation} (SE).

Among the exact solutions of the SE, we highlight the Rabi problem
\cite{Rabi2,Rabi}, which has applications in a wide variety of fields, such as quantum optics, quantum computing, condensed matter, molecular, atomic and particle physics.
Two-level systems can also be used as a model for open
systems, those which interact with the environment in which they are embedded.
Although the interaction problem is well-formulated in classical physics, it
is not yet fully comprehended at the quantum level. One of the reasons for the
lack of a proper quantum description of the interacting process is that open
systems are often described by non-Hermitian Hamiltonians \cite{Rot09}, and
consequently, by nonunitary theories. Due to the probabilistic interpretation
of quantum mechanics, the notion of a nonunitary theory raises important
questions. Despite that, nonunitary theories have drawn some attention in the
physics community through the study of a certain class of non-Hermitian
operators called \textit{pseudo-Hermitian operators} (PHOs). PHOs define the
so-called \textit{pseudo-Hermitian quantum mechanics} (PHQM). In PHQM, the
freedom in defining an inner-product in the physical Hilbert spaces is
explored to recover unitarity. Therefore, one may think that the notion of
nonunitarity arises because one is using the ``wrong'' inner product.

The freedom in choosing the inner product has already been studied
\cite{Dirac30,Pauli43,Gupta50,Bleuler50,Surdashan61,Lee69}. These early
developments attempted to recover unitarity from systems using what they
called \textit{indefinite-metrics quantum theories} (the terminology
``indefinite-metrics'' stands for
nonpositive-definite inner products). More recently, non-Hermitian
Hamiltonians with real eigenvalues were considered (see, for example,
Ref.~\cite{Scholtz92}). Later on, a series of papers
\cite{Bender98,Bender02,Bender03,Bender04,Bender07} exploring whether a
Hamiltonian $\hat{H}$ must be Hermitian or not were proposed. The authors argued that a
weaker and physically transparent condition for the reality of the spectrum of
$\hat{H}$ is the presence of $\mathcal{PT}$ symmetry, where $\mathcal{P}$
stands for the parity operator and $\mathcal{T}$ stands for the time-reversal
operator.%
\footnote{$\left\langle x,\mathcal{P}\psi\left(  t\right)
\right\rangle =\psi\left(  -x,t\right)  $ and $\left\langle x,\mathcal{T}%
\psi\left(  t\right)  \right\rangle =\bar{\psi}\left(  x,-t\right)  $, where
the bar denotes complex conjugation.}%
Also, it was shown that if $\hat{H}$ has
an unbroken $\mathcal{PT}$ symmetry, there is an operator $\mathcal{C}$,
commuting with $\hat{H}$, that allows one to define a positive-definite inner
product, with a metric operator given by $\eta=\mathcal{CPT}$.

The issue of what are the necessary and sufficient conditions for the reality
of the spectrum of a linear operator were explored in Refs.~\cite{Ali02-1,Ali02-2,Ali02-3,Ali04,Ali03}. It turns out that the answer to
this problem propelled the research in PHQM. It was shown that, albeit
relevant, the role played by the $\mathcal{PT}$ symmetry and the $\mathcal{C}$
operator is not a fundamental one. Indeed, it can be seen from PHQM that
$\eta=\mathcal{CPT}$ is just an example of a positive-definite metric operator
\cite{Ali10}. In fact, the existence of a preferred metric, and its physical
meaning, is an open issue in the PHQM. There are several contexts where
pseudo-Hermitian operators appear \cite{Ali10}. In special, recent treatments
of topological aspects of non-Hermitian systems use the framework of PHQM
\cite{Top1,Top2,Top3,Top4,Top5,Top6,Top7,Top8,Ana19,ZenYa20}.

A subtle point regarding quantization in general, and quantization in the PHQM
framework in particular, is that canonical transformations, which are
transformations on the level of the algebra of operators, do not necessarily
translate as isometries or unitary transformations between the Hilbert spaces
upon which these operators act \cite{And93}. When one is faced with
nonunitary canonical transformations, for instance, in the
infinite-dimensional case, a physical meaning for these transformations can be
established by looking at the classical limit of the theory
\cite{Ali04,Ali05,Ali10}. This procedure is called 
$\eta$-\textit{pseudo-Hermitian canonical quantization}.

For the present work, the important observation is that there is no usual
classical analog for a system with fermionic degrees of freedom. Nevertheless,
quantization schemes can still be defined in the context of
\textit{pseudoclassical mechanics} \cite{Cas76,Cas762,Ber77}, in which
Grassmann variables are used as phase-space coordinates. In this picture, the
Grassmanian degrees of freedom should be quantized with anticommutation
relations, rather than with commutation relations. The latter is of course a
well-known scheme for quantization of fermionic degrees of freedom, such as spin.

In this paper, the pseudo-Hermitian treatment will be extended to the
pseudoclassical framework. Despite the existing treatments concerning
pseudoclassical mechanics, its relation with pseudo-Hermitian theories was not
yet fully analyzed. The aim of this work is to exploit the latter at the level
of canonical transformations, considering both the pseudo-Hermitian quantum
theory and its pseudoclassical limit. For this purpose, complex external
fields, associated with nonunitary systems, will be considered. We then study
the classical correspondence in order to assign a physical meaning for the
complex fields. We construct a covariant quantization scheme which maps
canonically related pseudoclassical theories with real and complex external
fields to unitarily equivalent quantum realizations, such that there is a
unique metric-inducing isometry between the distinct Hilbert spaces. In this
setting, the pseudo-Hermiticity condition for the operators induces an
involution which guarantees the reality of the corresponding symbols, even in
the presence of complex external fields. We apply these developments to
propose a damped version of the Rabi problem, which could have important
implications in related areas. Furthermore, possible experimental tests for
the theory are proposed.

This work is organized as follows. In Sec.~\ref{Pseudo}, the basic
theoretical setup is established, with a revision of the notation used in the
present development. In Sec.~\ref{correspondence}, a classical-quantum
 correspondence is proposed and explored. A
physical realization of the proposed theoretical framework is constructed in
Sec.~\ref{realization}, where the Rabi problem is extended and its
generalization analyzed. Also in this section, we propose an experimental arrangement for the verification of the main theoretical results developed in the present work. In Sec.~\ref{final-remarks} final remarks and
future perspectives are presented. 
Further details on the pseudoclassical model considered and the quantization procedure employed are presented in Appendixes~\ref{appendix-pseudoclassical} and \ref{appendix-quantization}.
Units where $\hbar=1$ are used in this work, except where otherwise indicated.

\section{Pseudo-Hermitian and pseudoclassical frameworks}
\label{Pseudo}

\subsection{Pseudo-Hermitian theories}

Simply put, pseudo-Hermitian operators are operators which are not Hermitian
or symmetric with respect to the canonical or natural inner product, but which
are Hermitian with respect to some (positive-definite) inner
product.%
\footnote{We note that there is a broader definition of pseudo-Hermiticity where the product is not necessarily positive definite \cite{Ali02-4}. This characterization takes into account operators whose eigenvalues appear as complex-conjugate pairs. The restriction to the real spectrum is sometimes referred to as cryptoHermitian or quasi-Hermitian \cite{antoine2014}.}%
The treatment of
pseudo-Hermitian operators starts with the observation that non-Hermitian
matrices (that is, matrices that are not equal to their own conjugate
transpose) can have real eigenvalues. It follows that the spectra of the
related operators can be associated with physical observables in the quantum
description of a system. Taking a pseudo-Hermitian operator as the Hamiltonian
of the system, an evolution operator can be constructed in such way that the
time evolution is unitary \cite{Ali10}. This formalism is the base of the
pseudo-Hermitian quantum mechanics.

Pseudo-Hermitian operators in general will not have orthogonal eigenvectors
corresponding to distinct eigenvalues, as do Hermitian and normal operators.
Despite of this problem, the familiar probabilistic interpretation of quantum
mechanics can be recovered with a convenient choice of inner product.

Let us consider the pseudo-Hermitian formalism associated with the problem at
hand. Let $\mathcal{H}$ be a finite-dimensional Hilbert space isomorphic to
$\mathbb{C}^{n}$ with the canonical%
\footnote{The canonical inner product
$\langle.\,,.\rangle$ is defined as 
$\langle z,w\rangle=\bar{z}_{1} w_{1}+\cdots\bar{z}_{n}w_{n}$, 
where $z,w\in\mathbb{C}^{n}$.}%
inner product
$\langle\cdot\,,\cdot\rangle$, 
$\mathcal{H}\simeq(\mathbb{C}^{n},\langle\cdot\,,\cdot\rangle)$. 
We denote the adjoint of an operator $T$ with respect
to the canonical inner product to be $T^{\dagger}$.

Now let $\eta:\mathcal{H} \rightarrow\mathcal{H}$ and define
\begin{equation}
\left\langle x,y\right\rangle _{\eta}\equiv\left\langle x,\eta y\right\rangle
\,,\,\,\forall x,y\in\mathbb{C}^{n}\,. \label{etaproduct}%
\end{equation}
The sesquilinear form $\langle\cdot\,,\cdot\rangle_{\eta}$ is an inner product
in $\mathbb{C}^{n}$ if and only if
\begin{equation}
\eta=P^{\dagger}P \label{PP}%
\end{equation}
for some invertible $P$. Let us denote this new Hilbert space as
$\mathcal{H}_{\eta}\simeq(\mathbb{C}^{n},\langle\cdot\,,\cdot\rangle_{\eta})$.
We denote $\eta$ as the metric operator. In this case, an operator
$T:\mathcal{H}\rightarrow\mathcal{H}$ is pseudo-Hermitian or $\eta$-Hermitian
if and only if it is symmetric with respect to the inner
product~(\ref{etaproduct}). In other words, $T:\mathcal{H}\rightarrow
\mathcal{H}$ is pseudo-Hermitian if and only if it is Hermitian as an operator
on $\mathcal{H}_{\eta}$. It follows that an $\eta$-Hermitian operator $T$
satisfies
\begin{equation}
T=\eta^{-1}T^{\dagger}\eta\,. \label{PHd}%
\end{equation}
It should be noticed that the metric operator $\eta$ is not unique. In fact,
if $A$ is any invertible operator which commutes with the $\eta$-Hermitian
operator $T$, then $T$ is Hermitian with respect to the inner product
$\langle\cdot\,,\cdot\rangle_{\tilde{\eta}}$ with metric $\tilde{\eta
}=A^{\dagger}\eta A$.

For the specific case of the generic two-level system SE in Eq.~(\ref{SE}),
defined in terms of the Hamiltonian operator
\begin{equation}
\hat{H}=\frac{1}{2}\left(
\begin{array}
[c]{cc}%
B_{3} & B_{1}-\mathrm{i}B_{2}\\
B_{1}+\mathrm{i}B_{2} & -B_{3}%
\end{array}
\right)  \, ,
\end{equation}
with eigenvalues
\begin{equation}
E_{\pm}=\pm\frac{1}{2}\sqrt{B_{1}^{2}+B_{2}^{2}+B_{3}^{2}} \, , \label{spec}%
\end{equation}
one sees that the operator $\hat{H}$ is pseudo-Hermitian if and only if
\begin{equation}
\det(\hat{H})=-\frac{1}{4}\left(  B_{1}^{2}+B_{2}^{2}+B_{3}^{2}\right)
\in\mathbb{R}_{-}\,, \label{Cond1}%
\end{equation}
since this corresponds to real eigenvalues \cite{Ali02-4}.

As we will show in Sec.~\ref{quant}, a choice of metric $\eta$ induces an
isometry $\mathcal{M}$ between the Hilbert spaces $\mathcal{H}$ and
$\mathcal{H}_{\eta}$ such that the Hermitian operators on $\mathcal{H}$ are
mapped to Hermitian operators on $\mathcal{H}_{\eta}$. On the other hand,
these operators can be seen as images of quantization maps on pseudoclassical
phase spaces which are themselves related by canonical transformations. The
symbols of these operators, according to each quantization map, are real
functions in the respective pseudoclassical phase space.

\subsection{Pseudoclassical theories}
\label{pseudo-classical-theories}

Let us introduce the pseudoclassical framework used in this work. Further comments on this setup are presented in Appendix~\ref{appendix-pseudoclassical}.
Consider a Grassmann algebra $G_{3}(\xi)$ over the complex field
$\mathbb{C}$ with generators $\xi_{i},~i=1,2,3$, $\xi_{i}^{2}=0$, and the
pseudoclassical Lagrangian,
\begin{equation}
L=\frac{\mathrm{i}}{2}\xi_{i}\dot{\xi}_{i}-H\left(  \xi\right)  \,.
\label{lagrangian}
\end{equation}
By requiring that $\xi_{i}$ transform as a vector under $O\left(  3\right)  $,
it is natural to consider a rotational- and parity-invariant theory. In this
case, from the development presented in appendix~\ref{appendix-pseudoclassical}, the Hamiltonian $H$ must be of the form
\begin{equation}
H=H_{B}=-\frac{\mathrm{i}}{2}\varepsilon_{ijk}\xi_{i}\xi_{j}B_{k}%
\,,\label{HSC}%
\end{equation}
where $B_{k}$ transforms as a pseudovector (for instance, like the magnetic
field). Thus, the equation of motion for $\xi_{i}$ becomes%
\begin{equation}
\dot{\xi}_{i}=\{\xi_{i},H\}_{D(\phi)}=-\varepsilon_{ijk}\xi_{j}B_{k}%
\,.\label{CEM}%
\end{equation}
Where $\{\cdot,\cdot\}_{D(\phi)}$ is the Dirac brackets over the set of
second-class constraints
\begin{equation}
\phi_{i}=\pi_{i}-\frac{\mathrm{i}}{2}\xi_{i}\,,\label{constraints2}%
\end{equation}
and $\pi_{i}$ the conjugate momenta
\begin{equation}
\pi_{i}=\frac{\partial L}{\partial\dot{\xi}_{i}}\,.
\end{equation}
See Appendix~\ref{appendix-pseudoclassical} for details. We recognize~(\ref{CEM}) as the classical precession equation, like a magnetic moment immersed in a magnetic field $\mathbf{B}=\left(  B_{1},B_{2},B_{3}\right)$.

Of particular interest for the present work is the role of involution and
canonical transformations in the pseudoclassical formalism. For a general
function
\begin{equation}
f\left(  \xi\right)  =f_{0}+f_{i}\xi_{i}+f_{ij}\xi_{i}\xi_{j}+\frac
{\mathrm{i}}{3!}k_{f}\varepsilon_{ijk}\xi_{i}\xi_{j}\xi_{k}%
\,,\label{general-f2}%
\end{equation}
as in Eq.~(\ref{general-f}) in Appendix~\ref{appendix-pseudoclassical}, we define an involution 
$\ast :G_{3}(\xi)\rightarrow G_{3}(\xi)$ 
such that its action on the generators
$\xi_{i}$ is given by
\begin{equation}
\xi_{i}^{\ast}=\xi_{i},\,i=1,2,3\,.\label{star-involution}%
\end{equation}
Therefore, elements of the real subalgebra (those for which $f^{\ast}=f$) are
given by Eq.~(\ref{general-f2}) with $f_{0},f_{i},k_{f}\in\mathbb{R}$ and
$f_{ij}=\bar{f}_{ji}$. In particular, the $\ast$-involution as defined above
yields $H_{B}(\xi)$ in Eq.~(\ref{HSC}) to be real when $\mathbf{B}%
\in\mathbb{R}^{3}$. That is,%
\begin{equation}
\mathbf{B}\in\mathbb{R}^{3}\Longleftrightarrow H_{B}(\xi)=H_{B}^{\ast}(\xi)\,.
\end{equation}

Suppose we consider a linear canonical transformation on the pseudomechanical
phase space, defined as a map $(\xi,\pi)\mapsto(\zeta,\varpi)$, which
preserves the symplectic structure in that the only nonvanishing Poisson
brackets between the new coordinates are 
$\{\zeta_{i}(\xi,\pi),\varpi_{j}(\xi,\pi)\}=\delta_{ij}$. 
Due to the constraints $\phi$ in
Eq.~(\ref{constraints2}), we observe that $\pi$ is proportional to $\xi$, so
we write the linear canonical transformation simply as
\begin{equation}
\zeta_{i}=R_{ik}\xi_{k}~\text{and}~\varpi_{j}=R_{jl}\pi_{l}\,.\label{ctc}%
\end{equation}
Then, demanding that this transformation is canonical implies 
$RR^{T}=\mathbb{I}$, 
that is, $R$ is an orthogonal matrix. In principle, $R$ can have
complex entries, so $R\in O(3,\mathbb{C})$. Furthermore, under this
transformation the Hamiltonian function~(\ref{HSC}) becomes
\begin{equation}
H_{F}\left(  \zeta\right)  =-\frac{\mathrm{i}}{2}\varepsilon_{ijk}\zeta
_{i}\zeta_{j}F_{k}\,,\label{zeta-hamiltonian}%
\end{equation}
where
\begin{equation}
F_{k}=(\det R)R_{kl}B_{l} \,.
\label{FfromB}
\end{equation}
Relation~(\ref{FfromB}) implies that
\begin{equation}
F^{2}=F_{i}F_{i}=\left(  \det{R}\right)  ^{2}R_{ij}R_{ik}B_{j}B_{k}%
=\delta_{jk}B_{j}B_{k}=B^{2}.\label{Cond0}%
\end{equation}
Thus, if $\mathbf{B}$ is a real field, then from the previous relation it
follows that $F^{2}$ is a positive real number for an arbitrary complex field
$\mathbf{F}$.

Indeed, considering a complex field $\mathbf{F}$, one can define an involution
such that~(\ref{zeta-hamiltonian}) is real with respect to the new involution.
Initially, let us look at functions on the Grassmann algebra $G_{3}(\zeta)$
with generators $\{\zeta_{i}\}_{i=1}^{3}$, which are given by
\begin{equation}
g=g^{0}+g_{i}^{1}\zeta_{i}+g_{ij}^{2}\zeta_{i}\zeta_{j}+\mathrm{i}k_{g}%
\frac{1}{3!}\varepsilon_{ijk}\zeta_{i}\zeta_{j}\zeta_{k}\,. \label{general-g}%
\end{equation}
Then an involution $+:G_{3}(\zeta)\rightarrow G_{3}(\zeta)$ can be defined,
whose action on generators is given by
\begin{equation}
\zeta^{+}=\zeta^{\ast} \, , \label{plus-involution}%
\end{equation}
where the $\ast$-involution is presented in Eq.~(\ref{star-involution}) and
the $\zeta$-terms above are taken as function of $\xi$. As a result, the even
subalgebra of $G_{3}(\zeta)$ is given by the functions~(\ref{general-g}) with
$g^{0}\in\mathbb{R}$, $g^{1}=RR^{\dagger}\bar{g}^{1}$, $R^{T}g^{2}R=\left(
R^{T}g^{2}R\right)  ^{\dagger}$, and $k_{g}\in\mathbb{R}$. One can also show
that the even subalgebras of $G_{3}(\xi)$ and $G_{3}(\zeta)$ are isomorphic,
since $f=f^{\ast}\Leftrightarrow g=g^{+}$ where $f(\xi)=g(\zeta(\xi))$. It
follows that the Hamiltonian function $H_{F}\left(  \zeta\right)  $ in
Eq.~(\ref{zeta-hamiltonian}) is real with respect to the $+$~involution~(\ref{plus-involution}), that is,
\begin{equation}
\mathbf{B}\in\mathbb{R}^{3}\Longleftrightarrow H_{F}(\zeta)=H_{F}^{+}%
(\zeta)\,.
\end{equation}

\section{Classical-quantum correspondence}
\label{correspondence}

\subsection{Quantization and hermiticity}
\label{quant}

In this section we develop the quantization procedure employed in the present work. Complementary material is presented in Appendix~\ref{appendix-quantization}.
Analogously to the classical case~(\ref{ctc}),
let us consider the canonical transformation $\zeta_{i}=R_{ij}\xi_{j}$ with
$R\in O(3,\mathbb{C})$. 
As it is described in detail in Appendix~\ref{appendix-quantization}, on the
Grassmann algebras $G_{3}(\xi)$ and $G_{3}(\zeta)$ we can define 
quantization maps $Q:G_{3}(\xi)\rightarrow L(\mathcal{H})$ and 
$Q^{\prime}:G_{3}(\zeta)\rightarrow L(\mathcal{H}_{\eta})$. 
A natural question is then
what is the relation between $Q(f)$ and $Q^{\prime}(g)$, where 
$g(\zeta)=f(\xi(\zeta))$. To address this issue, let us take $P=\mathcal{M}^{-1}$ in
the expression~(\ref{PP}), $\eta=P^{\dagger}P$, so
\begin{equation}
\eta=(\mathcal{M}\mathcal{M}^{\dagger})^{-1}\,.\label{etaM}%
\end{equation}
Then, from Eq.~(\ref{etaproduct}), we see that $\mathcal{M}:\mathcal{H}%
\rightarrow\mathcal{H}_{\eta}$ is the isometry
\begin{equation}
\langle\phi,\psi\rangle=\langle\mathcal{M}\phi,\mathcal{M}\psi\rangle_{\eta
}\label{isometry}%
\end{equation}
for all $\phi,\psi\in\mathbb{C}^{2}$. Thus, for $\phi^{\prime}=\mathcal{M}%
\phi$ and $\psi^{\prime}=\mathcal{M}\psi$, one has
\begin{equation}
\langle\phi^{\prime},Q^{\prime}(g)\psi^{\prime}\rangle_{\eta}=\langle
\phi,\mathcal{M}^{-1}Q^{\prime}(g)\mathcal{M}\psi\rangle\,.
\end{equation}
Since $f$ and $g$ represent the same classical state (i.e., are related by a
canonical transformation), one has the familiar relation between the operators
of the corresponding functions:
\begin{equation}
Q^{\prime}(g)=\mathcal{M}Q(f)\mathcal{M}^{-1}\,.\label{similarity}%
\end{equation}

Moreover, let $Q^{\prime+}(g)$ denote the adjoint of $Q^{\prime}(g)$ in the
inner product $\langle\cdot\,,\cdot\rangle_{\eta}$ in Eq.~(\ref{etaproduct}).
It follows from the definition~(\ref{etaproduct}) that $Q^{\prime+}%
(g)=\eta^{-1}Q^{\prime\dagger}(g)\eta$. Using the results~(\ref{isometry}) and
(\ref{similarity}), it is obtained that $Q^{\prime+}(g)=\mathcal{M}Q^{\dagger
}(f)\mathcal{M}^{-1}$. Thus, for real $g$ [with respect to the $+$~involution presented in Eq.~(\ref{plus-involution})], the corresponding operator is
symmetric, $Q^{\prime+}(g)=Q^{\prime}(g)$, since real $g$ ($g^{+}=g$) implies
real $f$ ($f=f^{\ast}$), and $Q^{\dagger}(f)=Q(f)$. The similarity
relation~(\ref{similarity}) preserves the canonical relation [see
relation~(\ref{AC}) in Appendix~\ref{appendix-quantization}],
\begin{equation}
\lbrack Q(\xi_{i}),Q(\xi_{j})] = \delta_{ij}\,,
\label{AC2}%
\end{equation}
and can be regarded as a quantum canonical transformation induced by the
classical canonical transformation~(\ref{ctc}).

By means of the relation $\eta=(\mathcal{MM}^{\dagger})^{-1}$, we see that
$\eta\rightarrow\eta$ if $\mathcal{M}\rightarrow\mathcal{M}U$, for unitary
$U$, $U^{\dagger}=U^{-1}$. Let us call $Q_{U}^{\prime}$ the quantization map
with isometry $\mathcal{M}U$. Then, the relation between $Q^{\prime}$ in
Eq.~(\ref{similarity}) and $Q_{U}^{\prime}$ is $Q_{U}^{\prime}=S^{+}Q^{\prime
}S$ where $S=\mathcal{M}(\mathcal{M}U)^{-1}$. That is $Q_{U}^{\prime}$ is
$+$-unitarily equivalent\footnote{$S^{+}=\eta^{-1}S^{\dagger}\eta$ is the
adjoint with respect to the $\eta$-inner product.} to $Q^{\prime}$:
\begin{equation}
\langle\phi,Q_{U}^{\prime}\psi\rangle_{\eta}=\langle S\phi,Q^{\prime}%
S\psi\rangle_{\eta}\,,
\end{equation}
and
\begin{equation}
\langle S\phi,S\psi\rangle_{\eta}=\langle\phi,\psi\rangle_{\eta}\,.
\end{equation}

A unitary representation of the Clifford algebra~(\ref{AC2}) on 
$\mathbb{C}^{2}$ is given by the Pauli matrices $\sigma_{i}$ as
\begin{equation}
Q(\xi_{i})=\frac{\sigma_{i}}{\sqrt{2}} \,.
\end{equation}
Then, following the fermionic symmetric ordering [see relation~(\ref{symmetrization}] in Appendix~\ref{appendix-quantization}),
the Hamiltonian operator $\hat{H}_{B}\equiv Q(H_{B})$ [image of (\ref{HSC}) by
the quantization map $Q$] is
\begin{equation}
\hat{H}_{B}=\frac{1}{2} \, \boldsymbol{\sigma}\cdot\mathbf{B} \,.
\label{HB}%
\end{equation}
We recognize $\hat{H}_{B}$ as the Hamiltonian for the SE~(\ref{SE}). Given a realization of the algebra~(\ref{AC2}), it is immediate to write a
realization for the operators $Q(g(\zeta))$ using relation~(\ref{similarity}).
For the particular case of the Hamiltonian function $H_{F}$ in
Eq.~(\ref{zeta-hamiltonian}), one has
\begin{equation}
Q^{\prime}(H_{F})\equiv\hat{H}_{F}=\mathcal{M}\hat{H}_{B}\mathcal{M}^{-1}\,.
\end{equation}
Since the above relation is a similarity transformation, both operators
$\hat{H}_{F}$ and $\hat{H}_{B}$ have the same eigenvalues, so from this point
of view $\mathcal{M}$ is a mere change of basis in $\mathbb{C}^{2}$.

There is a unique realization of the $Q^{\prime}(\zeta)$ algebra, up to the
sign of $\det R$, such that the Hamiltonian operator in both quantizations
have the same form, and that realization is
\begin{equation}
Q^{\prime}\left(  \zeta_{k}\right)  =
\det R \, \frac{\sigma_{k}}{\sqrt{ 2}} \,.
\end{equation}
In other words, up to a sign, if the $Q^{\prime}$ quantization is realized in
the usual representation by Pauli matrices, $\hat{H}_{F}$ is given by the
operator
\begin{equation}
\hat{H}_{F}=\frac{1}{2} \, \boldsymbol{\sigma} \cdot \mathbf{F} \, .
\label{HF}%
\end{equation}
Thus, starting from this requirement, one fixes the isometry $\mathcal{M}$
that will give~(\ref{HF}) from (\ref{HB}), and because of the
result~(\ref{etaM}), the $\eta$-inner product is also fixed. As a result, the
$Q^{\prime}$-quantization of $H_{F}$ will give the operator~(\ref{HF}).
Furthermore, one sees from this procedure that the isometry $\mathcal{M}$ is
unique. In Sec.~\ref{csl} we will provide a systematic way of constructing
the isometry.

In conclusion, $\hat{H}_{B}$ describes a quantum theory of a spin system
interacting with a real field $\mathbf{B}$, such that $\hat{H}_{B}=\hat{H}%
_{B}^{\dagger}$. At the same time, $\hat{H}_{F}$ describes a quantum theory of
a spin system interacting with a complex field $\mathbf{F}$ (with
$\text{Im}(\mathbf{F})\neq0$), such that $\hat{H}_{F}=\hat{H}_{F}^{+}$. In
this sense, what we have achieved so far is to connect the description of a
nonrelativistic spinning particle under a real field $\mathbf{B}$ with
another one with a complex field $\mathbf{F}$, such that the respective
Hamiltonians are real under their classic involutions, while the corresponding
operators are symmetric (or Hermitian) with respect to the inner products of
the Hilbert spaces whereupon they act. Both fields are connected by the
complex canonical transformation $R$ by Eq.~(\ref{FfromB}) which implies the
important algebraic relation~(\ref{Cond0}).

An important remark following from Eq.~(\ref{Cond0}) should be stressed here.
The condition $F^{2}\in\mathbb{R}_{+}$ is exactly the condition~(\ref{Cond1})
that the quantum Hamiltonian needs to fulfill so that it is pseudo-Hermitian.
In other words, the existence of a real field $\mathbf{B}$, canonically
related to a field $\mathbf{F}$ with $\text{Im}(\mathbf{F})\neq0$, ensures the
reality of the spectrum of $\hat{H}_{F}$, according to Eq.~(\ref{spec}). This
result implies in the existence of a metric operator $\eta$ that renders
$\hat{H}_{F}$ Hermitian. Furthermore, the same canonical transformation
connects the two pseudoclassical models whose Hamiltonians are real with
respect to the corresponding involutions.

\subsection{\label{csl}Canonical limit and classical correspondence}

Unlike the usual description of pseudo-Hermitian theories, where the metric
operator is not unique, we have seen in Sec.~\ref{quant} that the metric
derived from the isometry~(\ref{isometry}) is actually unique. We present in
this section a schematic way to construct this metric operator, which we call
``the canonical limit.''

Besides giving the explicit form of the metric, the canonical limit also
furnishes a physical interpretation to our pseudo-Hermitian setup. Vectors
related by the isometry $\mathcal{M}$ describe the same physical system. In
other words, the Hilbert spaces $\mathcal{H}\simeq(\mathbb{C}^{2},\langle
\cdot\,,\cdot\rangle)$ and $\mathcal{H}_{\eta}\simeq(\mathbb{C}^{2}%
,\langle\cdot\,,\cdot\rangle_{\eta})$ represent two physically equivalent
quantum descriptions (quantizations) of the same classical model, with two
classical description that differ by a canonical transformation. Therefore, in
order to give correct measurable results, the states must be handled with the
appropriate metric.

Let us consider an orthonormal basis $\left\{  \phi_{\pm}\right\}  $ in
$\mathcal{H}$. So, the states $\phi_{\pm}\in\mathcal{H}$ were prepared (or
measured) by the observer associated with the canonical metric in his quantum
description. While the states
\begin{equation}
\phi_{\pm}^{\prime}=\mathcal{M}\phi_{\pm} \label{iso}%
\end{equation}
were prepared by an observer associated with the $\eta$ metric. The states
$\left\{  \phi_{\pm}^{\prime}\right\}  $ form a orthonormal basis of
$\mathcal{H}_{\eta}$. One observer does not agree about the orthogonality of
the states prepared by the other. Thus these observers are using different
measurement apparatus to construct the quantum description (of the same
system). The use of the canonical metric on the state $\phi_{\pm}^{\prime}$
(or the metric $\eta$ on $\phi_{\pm}$) is physically meaningless. In the
present work, the states whose probabilities must be calculated with the
$\eta$ metric are denoted by primes. The physical description by the observer
associated with the metric $\eta$ is compatible with the presence of an
(effective) complex field $\mathbf{F}$ and the observer associated with the
canonical metric measures a real field $\mathbf{B}$. In other words, we
distinguish the observables 
$\hat{H}_{F}:\mathcal{H}_{\eta}\rightarrow \mathcal{H}_{\eta}$ 
and $\hat{H}_{B}:\mathcal{H}\rightarrow\mathcal{H}$. 
For every operator $A$ acting on $\mathcal{H}$ there is an equivalent operator
$A^{\prime}=\mathcal{M}A\mathcal{M}^{-1}$ acting on $\mathcal{H}_{\eta}$.

The classical and quantum descriptions of both observers, especially their
notion of orthogonality, must coincide when 
$\operatorname{Im}\left(\mathbf{F}\right)  \rightarrow 0$. 
To achieve this requirement it is
necessary to choose $\phi_{\pm}^{\prime}$ and $\phi_{\pm}$ in Eq.~(\ref{iso})
to be, respectively, the eigenvectors of $\hat{H}_{F}=\hat{H}_{F}^{+}$ and
$\hat{H}_{B}=\hat{H}_{B}^{\dagger}$.

Let us consider with more attention the limit 
$\operatorname{Im}\left(\mathbf{F}\right)  \rightarrow 0$.
Following the proposal presented in Ref.~\cite{SL2009}, we postulate that there are three real dimensionless parameters $\{\alpha_{i}\}$ such that
\begin{equation}
\operatorname{Im}(F_{i}) = \alpha_{i} V_{i} \, .
\label{def-alpha}
\end{equation}
The parameters $\{\alpha_{i}\}$ measure how far the Hamiltonian $\hat{H}_{F}$ is from being canonically Hermitian.
Thus, we are interested in systems where the canonical
Hermiticity of $\hat{H}_{F}$ is broken continuously, namely, with a
well-defined limit $\alpha_{i} \rightarrow 0$. 
This is a reasonable requirement considering future applications in concrete physical scenarios, where the $\{\alpha_{i}\}$ are expressed in terms of measurable quantities.

Considering~(\ref{def-alpha}), $\hat{H}_{F}$
becomes Hermitian with respect to the canonical inner product, and both
theories (defined by $\hat{H}_{B}$ and $\hat{H}_{F}$) will differ at most by a
unitary transformation. To implement this requirement, for a given
$\mathbf{F}$, we choose the real field $\mathbf{B}$ such that
\begin{equation}
\lim_{\alpha_{i}\rightarrow0}\mathbf{F}=\lim_{\alpha_{i}\rightarrow
0}\mathbf{B}\in\mathbb{R}^{3}\,.\label{sl}%
\end{equation}
We stress that, due to relation~(\ref{FfromB}), the real-valued field $\mathbf{B}$ also depends on $\{ \alpha_{i} \}$.
Thus, in order to relate $\mathbf{F}$ and $\mathbf{B}$ in the regime of vanishing $\alpha_{i}$, the
limit $\alpha_{i} \rightarrow 0$ must be considered for both fields in Eq.~(\ref{sl}).
As
we will see in a future example, Eq.~(\ref{sl}) gives us a prescription such
that, when $\alpha_{i}\rightarrow0$,
\begin{equation}
\phi_{\pm}^{\prime}\rightarrow\phi_{\pm}\Longrightarrow\mathcal{M}%
\rightarrow\mathbb{I}\Longrightarrow\eta\rightarrow\mathbb{I}\,.
\end{equation}
In summary, the canonical limit is defined to be the prescription~(\ref{sl}),
together with the unique isometry which defines $\eta$ and relates the
eigenvectors of $\hat{H}_{F}$ and $\hat{H}_{B}$.

We turn now our attention to the classical correspondence of two quantum
theories: one with a non-Hermitian Hamiltonian, and another with a Hermitian
Hamiltonian. From now on we will assume that non-Hermitian operators are those
for which there is no inner product with respect to which they are Hermitian.

We construct the classical correspondence by taking mean values of operators.
The dynamical variables are real numbers that we expect to be related with the
measurable behavior of the system. As will see, for non-Hermitian Hamiltonians
this averaging procedure does not recover the classical equations of motion.
On the other hand, the classical equations of motion are recovered for
pseudo-Hermitian Hamiltonians.

To show the above statement, let us first consider the following
non-Hermitian Hamiltonian $\hat{H}$,
\begin{equation}
\hat{H}=\frac{1}{2}\boldsymbol{\sigma}\cdot\left[  \operatorname{Re}%
(\mathbf{F})+\mathrm{i}\operatorname{Im}(\mathbf{F})\right]  \, ,
\end{equation}
which is non-Hermitian by construction since its eigenvalues are not real. We
generally define the classical correspondence as the normalized mean value
(with the appropriate inner product) of the spin operators $\{\sigma_{i}\}$;
that is,
\begin{equation}
n_{i}\equiv\frac{\left\langle \psi,\sigma_{i}\psi\right\rangle }{\left\langle
\psi,\psi\right\rangle }\,,\ \mathbf{n}^{2}=1\,. \label{ni}%
\end{equation}
In the present case, because the Hamiltonian is non-Hermitian and there is no
suitable inner product, we used the canonical inner product. For $\psi$ a
solution of the time-dependent Schr\"{o}dinger equation, we have
\begin{equation}
\dot{n}_{i}=\frac{1}{\left\langle \psi,\psi\right\rangle }\left[
\mathrm{i}\left\langle \psi,\left(  \hat{H}^{\dagger}\sigma_{i}-\sigma_{i}%
\hat{H}\right)  \psi\right\rangle -n_{i}\frac{d}{dt}\left\langle \psi
,\psi\right\rangle \right]  \,,
\end{equation}
or, in a vector notation,
\begin{equation}
\mathbf{\dot{n}}=-\mathbf{n}\times\operatorname{Re}\left(  \mathbf{F}\right)
-\mathbf{n\times}\left[  \mathbf{n}\times\operatorname{Im}\left(
\mathbf{F}\right)  \right]  \,. \label{nSE}%
\end{equation}

It follows that, when the external field is real, that is, when
$\operatorname{Im}\left(  \mathbf{F}\right)  =0$, relation~(\ref{nSE})
coincides with Feynman's results in Ref.~\cite{Fey57}. Also, in this case the
result~(\ref{nSE}) reproduces the precession equation~(\ref{CEM}) of the
pseudoclassical theory. However, for 
$\operatorname{Im}\left(  \mathbf{F}\right)  \neq0$, 
Eq.~(\ref{nSE}) has an additional term that leads to damping
of the dynamics of $\mathbf{n}$. 

The damping term cannot be obtained classically from the Lagrangian~(\ref{lagrangian}) simply taking the external field to have imaginary entries from the very start. The reason is that real and complex fields provide the same equation of motion~(\ref{CEM}).

At this point, we should mention that, when dealing with a real field
$\mathbf{B}$, one has the usual physical interpretation for the SE~(\ref{SE}), that is, of a charged particle interacting with an
external magnetic field. However, when dealing with a complex field, this
notion does not hold. Therefore, in order to give a physical meaning for a
complex field, we can look at Eq.~(\ref{nSE}) as
\begin{equation}
\mathbf{\dot{n}}=-\mathbf{n}\times\mathbf{F}_{\mathrm{eff}} \, ,\ \text{with
}\mathbf{F}_{\mathrm{eff}}=\operatorname{Re}\left(  \mathbf{F}\right)
+\mathbf{n}\times\operatorname{Im}\left(  \mathbf{F}\right)  \, . \label{ef}%
\end{equation}
In Eq.~(\ref{ef}), $\mathbf{F}_{\mathrm{eff}}$ plays the role of an effective
field in the precession equation. Therefore, when there is damping, the system
interacts with the environment in such manner that all the resulting
combinations of external and internal fields produce an effective field, which
can be represented as a complex external field. In the following section we
will give a concrete example.

Consider now the case where $\hat{H}$ is pseudo-Hermitian and therefore
$F^{2}\in\mathbb{R}_{+}$. We will show that in this case the theory is
unitary, and there are no damping terms in the equations of motion. Let
$\left\langle \cdot,\cdot\right\rangle _{\eta}$ be the inner product with
respect to which $\hat{H}$ is Hermitian. Then the classical correspondence
gives
\begin{equation}
n_{i}(t)=\frac{\left\langle \psi,\sigma_{i}\psi\right\rangle _{\eta}%
}{\left\langle \psi,\psi\right\rangle _{\eta}}=\left\langle \psi,\sigma
_{i}\psi\right\rangle _{\eta}\,, \label{nn}%
\end{equation}
rather than Eq.~(\ref{ni}). In this case $\hat{H}^{+}=\hat{H}$ and we have
\begin{equation}
\dot{n}_{i}\left(  t\right)  =\mathrm{i}\left\langle \psi,\left[  \hat
{H},\sigma_{i}\right]  \psi\right\rangle _{\eta}=-\varepsilon_{ijk}%
n_{j}\left(  t\right)  F_{k}\,,
\end{equation}
or, in a vector notation,
\begin{equation}
\dot{\mathbf{n}}=-\mathbf{n}\times\mathbf{F}\,. \label{Y}%
\end{equation}
The previous equation corresponds to the pseudoclassical equations of
motion~(\ref{CEM}) even when the external field has an imaginary part. The
pseudoclassical equations of motion are recovered from the classical
correspondence with the identification $\mathbf{n}\rightarrow\boldsymbol{\zeta
}$.

\begin{figure}[h]%
\centering
\includegraphics[width=0.45\columnwidth]{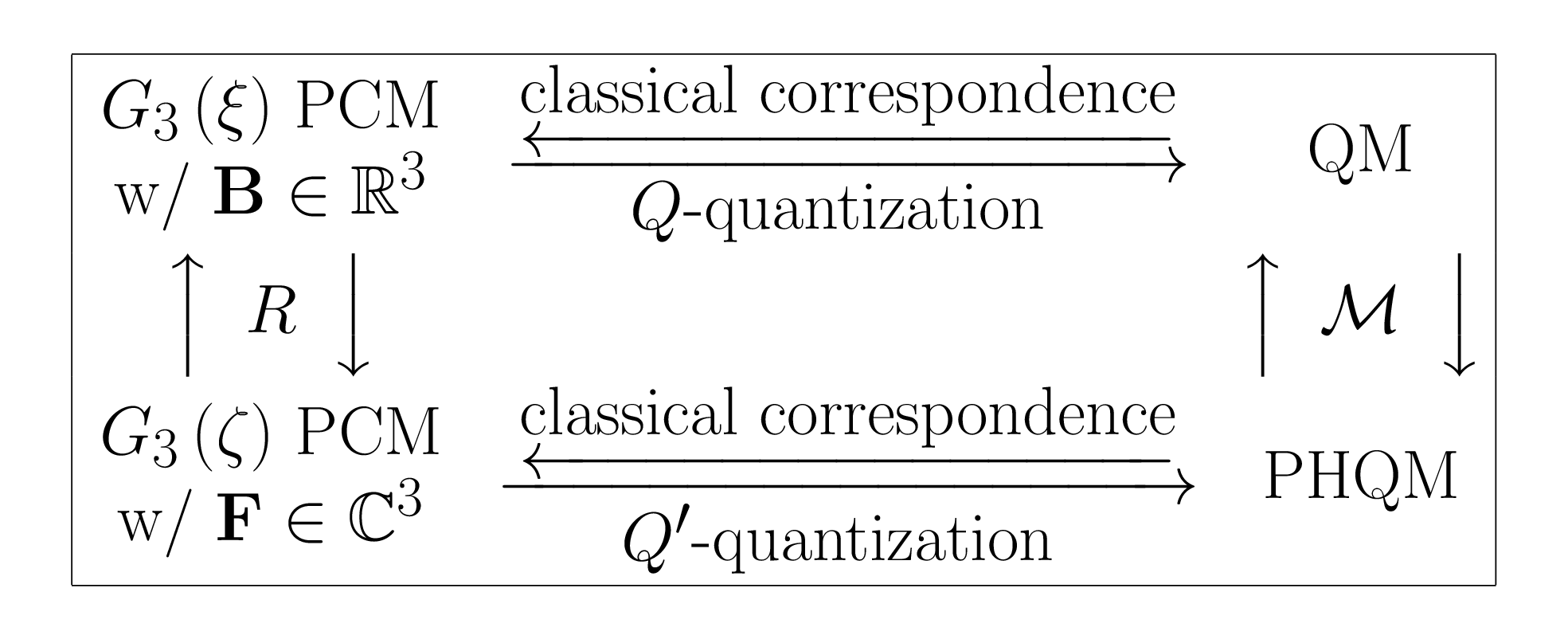}
\caption{Commutative diagram illustrating the classical-quantum correspondence
proposed. PCM denotes pseudoclassical mechanics, QM is short for the quantum
theory with Hamiltonian $Q(H_{B})$ and Hilbert space 
$\mathcal{H}\simeq(\mathbb{C}^{2},\langle\cdot\,,\cdot\rangle)$, 
while PHQM is short for the quantum theory with Hamiltonian $Q^{\prime}(H_{F})$ 
and Hilbert space
$\mathcal{H_{\eta}}\simeq(\mathbb{C}^{2},\langle\cdot\,,\cdot\rangle_{\eta})$.}
\label{diagram}%
\end{figure}

We conclude that a non-Hermitian Hamiltonian does indeed describe damping. On
the other hand, when the Hamiltonian is pseudo-Hermitian, the external field
fulfills the condition~(\ref{Cond0}) and the system does not present a
damping behavior. In particular, starting with a non-Hermitian Hamiltonian, we
can change the parameters of the effective field~(\ref{ef}) such that the
condition $F^{2}\in\mathbb{R}_{+}$ (with $\text{Im}(F)\neq0$) is satisfied. In
this case, there is a configuration of $\mathbf{F}$ such that the damping is
completely suppressed. In the following we use this property to propose a
possible measurable effect. For $F^{2}\in\mathbb{R}_{+}$, we can summarize the
results in the commutative diagram presented in Fig.~\ref{diagram}. We
emphasize that the classical-correspondence map in the diagram means that we
are able to formally obtain the pseudoclassical equations of motion after the
identification of $\mathbf{n}$ with corresponding Grassmann variable, either
$\boldsymbol{\xi}$ or $\boldsymbol{\zeta}$.

To consolidate the physical meaning to this correspondence, as well
as the physical interpretation of a complex field, let us introduce a concrete
scenario in the next section.

\section{Physical realization in the Rabi problem}

\label{realization}

\subsection{Preliminary results}

Now, we present some explicit examples and physical realizations for the
previous development, by considering the simplified case when
\begin{equation}
F_{2}=B_{2}=0\, . \label{CP}%
\end{equation}
As we will see, this particular restriction captures the essential points to
be studied in the present work.

Let us examine the following matrix
\begin{equation}
R=\frac{1}{B_{1}^{2}+B_{3}^{2}}\left(
\begin{array}
[c]{ccc}%
F_{1}B_{1}-B_{3}F_{3} & 0 & F_{1}B_{3}+B_{1}F_{3}\\
0 & -B_{1}^{2}-B_{3}^{2} & 0\\
F_{1}B_{3}+B_{1}F_{3} & 0 & -\left(  F_{1}B_{1}-B_{3}F_{3}\right)
\end{array}
\right)  \,. \label{T}%
\end{equation}
As one can explicitly check, $R\in SO(3,\mathbb{C})$ for arbitrary complex
vectors $\mathbf{F}$ and $\mathbf{B}$ is an explicit solution to the equation
$F_{k}=R_{kl}B_{l}$. In other words, $\det(R)=1$ and $R$ preserves the
symplectic structure~(\ref{DB}) commented in Appendix~\ref{appendix-pseudoclassical}. In the particular case~(\ref{T}), one has
additionally $R=R^{-1}$. Moreover, one can show that equation~(\ref{Cond0})
under the restriction~(\ref{CP}),
\begin{equation}
F_{1}^{2}+F_{3}^{2}=B_{1}^{2}+B_{3}^{2}\,, \label{Cond}%
\end{equation}
is a sufficient condition for the existence of $R$. As shown in
Sec.~\ref{pseudo-classical-theories}, for $\mathbf{B}\in\mathbb{R}^{3}$,
the Hamiltonians
\begin{equation}
H_{B}(\xi)=-\mathrm{i}\left(  B_{1}\xi_{2}\xi_{3}+B_{3}\xi_{1}\xi_{3}\right)
~\text{and}~H_{F}(\zeta)=-\mathrm{i}\left(  F_{1}\zeta_{2}\zeta_{3}+F_{3}%
\zeta_{1}\zeta_{2}\right)  ~ \label{HH}%
\end{equation}
are real in the sense of the involutions,
\begin{equation}
H_{B}(\xi)=H_{B}^{\ast}\left(  \xi\right)  ~\text{and}~H_{F}(\zeta)=H_{F}%
^{+}\left(  \zeta\right)  \, .
\end{equation}

Following our prescription for the canonical limit, we now use the
eigenvectors of $\hat{H}_{B}$ and $\hat{H}_{F}$ in order to construct the
metric operator $\eta$. Maintaining the convention of using primes to indicate
the states whose probabilities must be calculated with the $\eta$ metric, we
write the eigenvectors $\phi_{\pm}^{\prime}$\ of$~\hat{H}_{F}$, with
eigenvalues ${E_{F}}_{\pm}$, as
\begin{equation}
\phi_{\pm}^{\prime}=\frac{1}{F_{1}}\left(
\begin{array}
[c]{c}%
F_{3}\pm E_{F}\\
F_{1}%
\end{array}
\right)  \text{~},\ {E_{F}}_{\pm}=\pm\frac{E_{F}}{2}=\pm\frac{1}{2}\sqrt
{F_{1}^{2}+F_{3}^{2}}\,, \label{PsiH}%
\end{equation}
and the eigenvector $\phi_{\pm}$\ of $\hat{H}_{B}$, with eigenvalues~${E_{B}%
}_{\pm}$, as
\begin{equation}
\phi_{\pm}=\frac{1}{B_{1}}\left(
\begin{array}
[c]{c}%
B_{3}\pm E_{B}\\
B_{1}%
\end{array}
\right)  \,,\,{E_{B}}_{\pm}=\pm\frac{E_{B}}{2}=\pm\frac{1}{2}\sqrt{B_{1}%
^{2}+B_{3}^{2}}\,. \label{PsiH-B}%
\end{equation}
From Eq.~(\ref{Cond}) we see that $E_{F}=E_{B}\equiv E$. The isometry can be
read off from relation~(\ref{iso}) for the eigenvector $\phi_{\pm}$ and
$\phi_{\pm}^{\prime}$,%
\begin{equation}
\mathcal{M}=\frac{1}{F_{1}}\left(
\begin{array}
[c]{cc}%
B_{1} & F_{3}-B_{3}\\
0 & F_{1}%
\end{array}
\right)  \,,
\end{equation}
and the metric operator~(\ref{etaM})\ in $\mathcal{H}_{\eta}$ will be given
by
\begin{equation}
\eta=\frac{1}{B_{1}^{2}}\left(
\begin{array}
[c]{cc}%
\left\vert F_{1}\right\vert ^{2} & \bar{F}_{1}\left(  B_{3}-F_{3}\right) \\
F_{1}\left(  B_{3}-\bar{F}_{3}\right)  & B_{1}^{2}+\left\vert B_{3}%
-F_{3}\right\vert ^{2}%
\end{array}
\right)  \,. \label{Eta}%
\end{equation}
As expected from the general theory, one has the Hermiticity conditions
$\hat{H}_{B}=\hat{H}_{B}^{\dagger}$ and $\hat{H}_{F}=\hat{H}_{F}^{+}$.
Besides, by the canonical limit, if $\mathbf{B}=\mathbf{F}$ we have
$\mathcal{M}=\mathbb{I}$, $\eta=\mathbb{I}$, $\hat{H}_{B}=\hat{H}_{F}$.

Assuming that the operator $\hat{H}_{F}$ is time-independent, the dynamics is
simply obtained by exponentiation of $\hat{H}_{F}$. For instance, if one
wishes to evaluate a transition amplitude between the eigenvectors of
$\sigma_{3}\psi_{\pm}=\pm\psi_{\pm}$, that is, the states of ``spin-up'' $\psi_{+}$ and ``spin-down'' $\psi_{-}$\ in $\mathcal{H}$, we can construct
the corresponding states in $\mathcal{H}_{\eta}$ using the isometry
$\mathcal{M}$. This transition amplitude can be written as%
\begin{equation}
\left\langle \psi_{+}^{\prime},\psi^{\prime}\left(  t\right)  \right\rangle
_{\eta}=\left\langle \psi_{+}^{\prime},\exp\left(  -\mathrm{i}\hat{H}%
_{F}t\right)  \psi_{-}^{\prime}\right\rangle _{\eta}=-\mathrm{i}\frac{B_{1}%
}{E}\sin\left(  \frac{E}{2}t\right)  \,, \label{A}%
\end{equation}
where $\psi_{\pm}^{\prime}=\mathcal{M}\psi_{\pm}$. We note the oscillatory
behavior of (\ref{A}), which is a characteristic property of unitary theories.

Let us illustrate the above with the example of the Rabi oscillations in an
assumed damped two-level system \cite{Yac04}
\begin{equation}
F_{1}=V\in\mathbb{R}\,,\,F_{2}=0\,,\,F_{3}=\mathrm{i}\alpha\,,\,\alpha
^{2}<V^{2}\,. \label{fe}%
\end{equation}
The $\mathbf{F}$ field can be obtained from the canonical transformation
(\ref{T}) starting from any one of the following real $\mathbf{B}$ fields and
rotations thereof
\begin{align}
B_{1}  &  =B_{2}=0~,\ B_{3}=\pm\sqrt{V^{2}-\alpha^{2}}\,,\nonumber\\
B_{1}  &  =\pm\sqrt{V^{2}-\alpha^{2}}~,\ B_{2}=B_{3}=0\,.
\end{align}
However, the canonical limit~(\ref{sl}) implies the specific choice
\begin{equation}
B_{1}=\mathrm{sgn}\left(  V\right)  \sqrt{V^{2}-\alpha^{2}}\,,\ B_{2}%
=B_{3}=0\,. \label{ec}%
\end{equation}
As one can directly check, the Hamiltonian $\hat{H}_{F}$ is Hermitian
according to the metric~(\ref{Eta}), that is, it satisfies $\hat{H}_{F}%
=\eta^{-1}\hat{H}_{F}^{\dagger}\eta$. Given the configuration for $\mathbf{B}$
in (\ref{ec}), one can verify the canonical limit~(\ref{sl}) $\lim
_{\alpha\rightarrow0}\eta=\mathbb{I}$. The transition amplitude~(\ref{A})
between spin-up and spin-down states reads
\begin{equation}
\left\langle \psi_{+}^{\prime},\exp\left(  -\mathrm{i}\hat{H}_{F}t\right)
\psi_{-}^{\prime}\right\rangle _{\eta}=-\mathrm{i}\left[  \mathrm{sgn}\left(
V\right)  \right]  \sin\left(  \frac{\sqrt{V^{2}-\alpha^{2}}}{2}t\right)  \,.
\label{YAY}%
\end{equation}

Apart from the factor $\nicefrac{1}{2}$, due to our particular choice of
constants (see footnote~\ref{f1}), the oscillation frequency of the
amplitude~(\ref{YAY}) agrees with the one in Ref.~\cite{Yac04}. However, unlike in
Ref.~\cite{Yac04}, here the evolution is unitary and states do not lose their
normalization condition under time evolution. In general, there is a critical
value $\alpha_{c}$ of $\alpha$ for which 
 $\operatorname{Im}(E) = (V^{2}-\alpha^{2})^{\nicefrac{1}{2}} \neq 0$ 
if $\alpha>\alpha_{c}$. In the illustrative example
presented in this section, this critical value $\alpha_{c}=V$ can be read
from Eq.~(\ref{fe}). In some descriptions, the value $\alpha_{c}$ can be
 associated with a possible phase transition \cite{SL2009}. In this article, conditions~(\ref{Cond}) and 
$E_{F}=E_{B} \in \mathbb{R}$ are assumed.

\subsection{\label{ex}Rabi problem and the Gilbert damping term}

Let us now consider the more elaborate Rabi problem \cite{Rabi2,Rabi}. This is
a two-level system, consisting of a single electron fixed in the space, in
interaction with an external magnetic field given by
\begin{equation}
\mathbf{B}_{R}=\left(  B\cos\left(  \omega t\right)  ,B\sin\left(  \omega
t\right)  ,B_{z}\right)  \,, \label{RF}%
\end{equation}
with $B$, $B_{z}$ and $\omega$ real constants. We can eliminate the second
component of the $\mathbf{B}$ field by changing to a rotating reference frame
with the help of the rotation%
\begin{equation}
R_{z}(\omega t)=\exp\left(  \frac{\mathrm{i}\omega\sigma_{3}t}{2}\right)  \,.
\label{Rotacao}%
\end{equation}
In this rotating reference frame we have
\begin{equation}
B_{1}=B\,,\,B_{2}=0,~B_{3}=\delta\,,\,\delta=B_{z}-\omega\,, \label{BRR}%
\end{equation}
and time-independent Hamiltonian
\begin{equation}
\hat{H}_{R}=\frac{1}{2}\left(  \delta\sigma_{3}+B\sigma_{1}\right)  \,.
\label{Transformacao}%
\end{equation}
The transition amplitude between spin-up and spin-down states ($\sigma_{3}%
\psi_{\pm}=\pm\psi_{\pm}$) is given by the Rabi oscillations
\begin{equation}
\left\langle \psi_{+},\exp\left(  -\mathrm{i}\hat{H}_{R}t\right)  \psi
_{-}\right\rangle =-\mathrm{i}\frac{B}{\Omega_{R}}\sin\left(  \frac{\Omega
_{R}}{2}t\right)  \,,\ \Omega_{R}^{2}=B^{2}+\delta^{2}\,. \label{RO}%
\end{equation}
The $\delta$ factor is called detuning, while $\Omega_{R}$ and $\omega=B_{z}$
denote the Rabi frequency and resonance frequency respectively.

As we have seen in Sec.~\ref{csl}, a damped precession is characteristic of
a nonunitary evolution. Indeed, we can see that the damping term in
Eq.~(\ref{nSE}) arises exactly from the imaginary part of field, which is what
breaks the hermiticity of the Hamiltonian. Therefore, one can consider a
damped version of the Rabi problem by introducing an imaginary term in the
field~(\ref{BRR}). For this reason, we choose the external field to be
\begin{equation}
F_{1}=\frac{1+\mathrm{i}\alpha}{1+\alpha^{2}}B \, , \, F_{2}=0 \, ,\, F_{3}
=\frac{1+\mathrm{i}\alpha}{1+\alpha^{2}}B_{z}-\omega\,,\ \alpha\in
\mathbb{R}\,. \label{f}%
\end{equation}
In the limit $\alpha\rightarrow0$, this field configuration reduces to the
original Rabi problem characterized by~(\ref{BRR}) in the rotating frame. For
arbitrary values of the parameters $B$, $B_{z}$, $\omega$, $\alpha$, the
Hamiltonian $\hat{H}_{F}$ is non-Hermitian, resulting in a damped behavior.

The time-dependent field configuration for the damped Rabi setup in the
nonrotating frame is
\begin{equation}
\mathbf{F}_{R}=\left(  F_{1}\cos\left(  \omega t\right)  ,F_{1}\sin\left(
\omega t\right)  ,F_{3}\right)  =\frac{1+\mathrm{i}\alpha}{1+\alpha^{2}%
}\mathbf{B}_{R}\,, \label{Fr}%
\end{equation}
which reduces to the original Rabi problem described by $\mathbf{B}_{R}$
in~(\ref{RF}) when $\alpha$ is set to zero. We can now obtain the classical
correspondence, which we interpret as the behavior of the damped system as
actually measured. Substituting the field configuration~(\ref{Fr}) in
Eq.~(\ref{nSE}), we have
\begin{equation}
\mathbf{\dot{n}}=-\frac{1}{1+\alpha^{2}}\mathbf{n}\times\mathbf{B}_{R}%
-\frac{\alpha}{1+\alpha^{2}}\mathbf{n\times}\left(  \mathbf{n}\times
\mathbf{B}_{R}\right)  \,. \label{LLGEquation}%
\end{equation}

A physical interpretation can now be provided for the parameter $\alpha$. The
above equation describes a damped precession of the magnetic moment. As is
well known, this phenomenon can be adequately described by the
\textit{Landau-Lifshitz-Gilbert} (LLG) equation \cite{Gil04}, which consists
of introducing an \textit{ad hoc} term in the undamped equation of motion. The
LLG equation, for the unit magnetization $\mathbf{\hat{n}}$, subject to a
magnetic field $\mathbf{B}$, has the form \cite{Lac11}
\begin{equation}
\frac{d\mathbf{\hat{n}}}{dt}=-\frac{1}{1+\alpha^{2}}\mathbf{\hat{n}}%
\times\mathbf{B}-\frac{\alpha}{1+\alpha^{2}}\mathbf{\hat{n}}\times\left(
\mathbf{\hat{n}}\times\mathbf{B}\right)  \,, \label{LLG}%
\end{equation}
where $\alpha$ is the Gilbert damping parameter. By comparing the LLG
equation~(\ref{LLG}) with the relation~(\ref{LLGEquation}) obtained via
classical correspondence, we see that the $\alpha$ parameter introduced in
Eq.~(\ref{f}) can be identified with the Gilbert damping parameter.

Even though we have just addressed the damped Rabi problem, the identification
of $\alpha$ with the Gilbert damping term is valid for a general effective
field $\mathbf{F}$ in the form
\begin{equation}
\mathbf{F}=\frac{1+\mathrm{i}\alpha}{1+\alpha^{2}}\mathbf{B}\,, \label{Field}%
\end{equation}
for any $\mathbf{B}\in\mathbb{R}^{3}$. This follows from the fact that the
classic correspondence equation~(\ref{nSE}) is exactly the LLG equation for
the field configuration~(\ref{Field}).

\subsection{The pseudo-Hermitian version of the Rabi problem}

In this section we choose the parameters $B,B_{z},\omega$ and $\alpha$ such
that the restriction~(\ref{Cond}) is satisfied, so that $\hat{H}_{F}$ is
(pseudo) Hermitian, $\hat{H}_{F}^{+}=\hat{H}_{F}$. We introduce the notation
\begin{equation}
F\equiv F_{1}=\frac{1+\mathrm{i}\alpha}{1+\alpha^{2}}B\,\,\,\text{and}%
\,\,\,\,\Delta\equiv F_{3}=\frac{1+\mathrm{i}\alpha}{1+\alpha^{2}}B_{z}%
-\omega\,, \label{RP}%
\end{equation}
to label the field components satisfying the condition~(\ref{Cond}). Now the
classical Hamiltonian
\begin{equation}
H_{F}=-\mathrm{i}\left(  F\zeta_{2}\zeta_{3}+\Delta\zeta_{1}\zeta_{2}\right)
\,, \label{RD}%
\end{equation}
is real ($H_{F}^{+}=H_{F}$). The specific choice of parameters can be found
from Eq.~(\ref{Cond1}), i.e., from $\operatorname{Im}(F^{2}+\Delta^{2})=0$,
\begin{equation}
B^{2}+\delta^{2}-\alpha^{2}\omega^{2}+\delta\omega\left(  1-\alpha^{2}\right)
=0\,. \label{Cond3}%
\end{equation}
It follows that
\begin{equation}
F^{2}+\Delta^{2}=-\delta\omega=B_{1}^{2}+B_{3}^{2}\,. \label{F2}%
\end{equation}

Even though $\mathbf{B}$ has not yet been determined, the
eigenvalues~(\ref{PsiH-B}) of $\hat{H}_{B}$ are known, because of
relation~(\ref{Cond}). The Hamiltonian $\hat{H}_{F}$ has the eigenvectors
$\phi_{\pm}^{\prime}$ and eigenvalues $E_{\pm}$:
\begin{equation}
\phi_{\pm}^{\prime}=\frac{1}{F}\left(
\begin{array}
[c]{c}%
\Delta\pm\Omega\\
F
\end{array}
\right)  \,,\,E_{\pm}=\pm\frac{\Omega}{2}\,,\,\Omega^{2}=F^{2}+\Delta^{2}\,.
\label{aa}%
\end{equation}
From Eq.~(\ref{F2}), for $\delta\omega>0$ the eigenvalues are purely
imaginary, however we only consider the case where $\delta\omega<0$, that is,
the case of real eigenvalues. Considering that the limit $\alpha\rightarrow0$
implies
\begin{equation}
\Delta\rightarrow\delta,~F\rightarrow B,~\Omega\rightarrow\Omega_{R}\,,
\end{equation}
we use the canonical limit to construct the eigenvectors $\phi_{\pm}$ of
$\hat{H}_{B}$, which has the same eigenvalues $E_{\pm}$:
\begin{equation}
\phi_{\pm}=\frac{1}{B}\left(
\begin{array}
[c]{c}%
\delta\pm\Omega_{R}\\
B
\end{array}
\right)  \,. \label{ab}%
\end{equation}
After calculating the eigenvectors in~(\ref{aa}) and (\ref{ab}),
one can determine the isometry $\mathcal{M}$,
\begin{equation}
\mathcal{M}=\frac{1}{F\Omega_{R}}\left(
\begin{array}
[c]{cc}%
B\Omega & \Delta\Omega_{R}-\delta\Omega\\
0 & F\Omega_{R}%
\end{array}
\right)  \,,
\end{equation}
and the metric operator $\eta$,
\begin{equation}
\eta=\frac{1}{B^{2}\Omega^{2}}\left(
\begin{array}
[c]{cc}%
\left\vert F\right\vert ^{2}\Omega_{R}^{2} & \bar{F}\Omega_{R}\left(
\delta\Omega-\Delta\Omega_{R}\right) \\
F\Omega_{R}\left(  \delta\Omega-\bar{\Delta}\Omega_{R}\right)  & B^{2}%
\Omega^{2}+\left\vert \delta\Omega-\Delta\Omega_{R}\right\vert ^{2}%
\end{array}
\right)  \,. \label{E}%
\end{equation}
The expression for $\eta$ in~(\ref{E}) satisfies the canonical limit
$\mathcal{\eta}\rightarrow\mathbb{I}$ when $\alpha\rightarrow0$. The explicit
form of $\hat{H}_{B}$ can be obtained from result~(\ref{similarity}), i.e.,
$\hat{H}_{B}=\mathcal{M}^{-1}\hat{H}_{F}\mathcal{M}$. Moreover, one can
determine the $\mathbf{B}$ field,
\begin{equation}
\mathbf{B}=\frac{\Omega}{\Omega_{R}}\left(  B,0,\delta\right)  \,, \label{B}%
\end{equation}
and the canonical transformation $R$ from Eq.~(\ref{T}).

To obtain the pseudo-Hermitian version of the damped Rabi problem in
the original (nonrotating) frame, one must rotate back the reference frame
with the rotation $R_{z}^{\prime}=\mathcal{M}R_{z}\mathcal{M}^{-1}$, where
$R_{z}\left(  -\omega\right)  $ is given in Eq.~(\ref{Rotacao}), that is,
\begin{equation}
\hat{H}_{F}^{\prime}=\mathrm{i}\frac{\partial R_{z}^{\prime}}{\partial
t}\left(  R_{z}^{\prime}\right)  ^{-1}+R_{z}^{\prime}\hat{H}_{F}\left(
R_{z}^{\prime}\right)  ^{-1}=\mathcal{M}\hat{H}_{B}^{\prime}\mathcal{M}^{-1}
\, ,
\end{equation}
with%
\begin{equation}
\hat{H}_{B}^{\prime}=\frac{1}{2\Omega_{R}}\left(
\begin{array}
[c]{cc}%
\delta\Omega+\omega\Omega_{R} & B\Omega\exp\left(  -i\omega t\right) \\
B\Omega\exp\left(  i\omega t\right)  & -\left(  \delta\Omega+\omega\Omega
_{R}\right)
\end{array}
\right)  \,.
\end{equation}
As expected, $\hat{H}_{B}^{\prime}$ is the field obtained from~(\ref{B}) by
the usual rotation~(\ref{Rotacao}). The Hamiltonian $\hat{H}_{F}^{\prime}$
keeps its pseudo-Hermiticity. In the canonical limit, not only we verify that
$\hat{H}_{F}^{\prime}\rightarrow\hat{H}_{B}^{\prime}$, but we also recover the
Hamiltonian associated with the Rabi problem in the nonrotating frame~(\ref{RF}).

Let us consider the dynamics of this model. Using Eq.~(\ref{E}) we can
determine the transition amplitude~(\ref{A}) between spin-up and spin-down
states,
\begin{equation}
\left\langle \psi_{+}^{\prime},\exp\left(  -\mathrm{i}\hat{H}_{F}t\right)
\psi_{-}^{\prime}\right\rangle _{\eta}=-\mathrm{i}\frac{B}{\Omega_{R}}%
\sin\left(  \frac{\Omega}{2}t\right)  \, . \label{P}%
\end{equation}
The frequency $\omega=B_{z}$ ($\delta=0\Rightarrow\Omega=0$) represents a
critical point, which can be associated with symmetry breaking. From
relations~(\ref{F2}) and (\ref{Cond3}) we have%
\begin{equation}
\Omega^{2}=%
\begin{cases}
\Omega_{R}^{2}+\frac{\alpha^{2}}{1-\alpha^{2}}\left(  \Omega_{R}^{2}%
-\omega^{2}\right)  \text{ for }\alpha\neq\pm1\\
\left\vert \delta\Omega_{R}\right\vert \text{ for }\alpha=\pm1
\end{cases}
\,.
\end{equation}
In summary, when condition~(\ref{Cond}) holds, the theory is unitary, and
there is no damping term in the equations of motion.

Previous results can furnish possible measurable effects. The main point is
that $\hat{H}_{F}$ is non-Hermitian, and thus there would be a damping term in
the equations of motion for any value of the external field, except if
(\ref{Cond3}) is valid. When condition~(\ref{Cond3}) is satisfied, the damping
effect disappears and the evolution of the system becomes unitary. From
Eq.~(\ref{F2}) we see that, when $\omega>0$, the pseudo-Hermitian regime can
only be reached for $\delta<0$. This means that it is not possible to suppress
damping with a frequency below the resonance frequency of the usual Rabi
problem. In addition, we can use Eq.~(\ref{Cond3}) to determine, for example,
$B$ as a function of the other parameters:
\begin{equation}
B^{2}=B_{z}\left[  \omega\left(  1+\alpha^{2}\right)  -B_{z}\right]  \text{
for~}B, \, B_{z}, \, \alpha \neq 0 \, . 
\label{Condition}%
\end{equation}
We interpret the condition~(\ref{Condition}) as the configuration of the
$\mathbf{B}$ field which injects energy in the system at the same rate the
system dissipates energy. In this case, the damping effect is completely
suppressed and the classical limit is again a precession movement described
by~(\ref{Y}), and not by the LLG equation~(\ref{LLG}).

\subsection{Experimental viability} 
\label{experiment}

We now consider the experimental implementation of the ideas introduced. In
the arrangement presented in the previous section, the suppression of
damping may be identified with the so-called steady-state precession or
self-sustained precessional motion (SSPM) \cite{EvgTsZ2016}, where an
effective field cancels the spin damping, generating a constant angle
precession. Although the direct measurement of the local magnetization is a
challenge in an actual laboratory experiment, the spin dynamics can be
controlled and determined by a current-induced magnetization. The
magnetization itself is controlled by a combination of the external field and
the current electrons spin-transfer-induced precession of magnetization.

By varying the applied magnetic field and the direct current irradiated with
high-frequency microwaves, it is possible to control the precession frequency,
obtaining values from the order of $1GHz$ up to $100GHz$ or larger. In fact,
the irradiation induces microwave currents, which are equivalent to the
arrangement where currents are fed to the contact through electric leads
\cite{7-89}.

The high-frequency dynamics of the magnetization can be measured directly by
detecting high-frequency voltage oscillations \cite{7-94}. This
current-induced magnetization is different from the setup we have considered,
since it is related to spin-transfer torque. However, we are interested in the
macrospin model with uniform magnetization. In this case, a description of the
magnetization dynamics using a modified LLG equation that can be written in
the form~(\ref{ef}) is largely equivalent \cite{xiao2005}.

A concrete example where the SSPM can be detected are the so-called spin
valves. A simple spin valve is formed by two conducting magnetic
materials, the free layer and the reference layer, separated by a (metallic or
insulator) spacer. The electric resistance of the device can be controlled
through the relative alignment of the magnetization in the layers. The spins
of the electron current are adjusted by a hard and thick magnetic layer called
the reference layer, while the layer where magnetization will be manipulated
is called the free layer. A typical setup uses a pinned antiferromagnet
reference layer coupled to a ferromagnetic free layer.

In this device an applied current generates a spin torque that, under specific
conditions, can cancel the Gilbert damping term in the LLG equation resulting inan undamped precession~(\ref{Y}). Namely, for the spin valve, the modified
LLG equation assumes the form \cite{EvgTsZ2016}
\begin{equation}
\frac{d\mathbf{\hat{n}}}{dt}=-\frac{1}{1+\alpha^{2}}\mathbf{\hat{n}}%
\times\mathbf{B}-\frac{\alpha}{1+\alpha^{2}}\mathbf{\hat{n}}\times\left(
\mathbf{\hat{n}}\times\mathbf{B}\right)  +a\mathbf{\hat{n}}\times\left(
\mathbf{\hat{n}}\times\mathbf{P}\right)  \,,\label{8.33}%
\end{equation}
where $\mathbf{P}$ is a fixed unit vector in the magnetization
direction of the pinned layer in the spin valve and the torque coefficient $a$ depends on
the applied current and may also depend on the angle between the magnetization
directions of the pinned and free layers \cite{8-70}. The effective field
$\mathbf{B}$ in Eq.~(\ref{8.33}) is the combination of the exchange field, the
anisotropy field, the demagnetization field, and the applied external field. A
qualitative and quantitative discussion of Eq.~(\ref{8.33}) can be found in
Refs.~\cite{EvgTsZ2016,xiao2005} and references therein.

It is possible to show \cite{Lac11} that the effect of the new term in the
modified LLG equation~(\ref{8.33}) is simply to change the effective
magnetic field to
\begin{equation}
\mathbf{B}=\left(  B_{x},B_{y},B_{z}\right)  
\longrightarrow
\left(  B_{x}
-\mathrm{i}aP_{x},B_{y}+\mathrm{i}aP_{y},B_{z}+\mathrm{i}aP_{z}\right)  \, .
\end{equation}
That is, the effect of the spin-transfer-induced is  equivalent to the addition of an imaginary term to the magnetic field already present in the original LLG equation. Therefore, by choosing the magnetization direction of the reference layer in the $z$-direction, that is $\mathbf{P}=(0,0,1)$, 
the spin-transfer-induced effect can be taken into account with the change
\begin{equation}
B_{z}\longrightarrow B_{z} + \mathrm{i} a \, .
\end{equation}
Repeating our previous development using the new field $B_{z}$, it is straightforward to show that condition~(\ref{Condition}) becomes
\begin{equation}
B^{2}=\omega\left(  1+\alpha^{2}\right)  B_{z}-B_{z}^{2}+\frac{a}{\alpha
}\left[  \alpha a-B_{z}\left(  1-\alpha^{2}\right)  +\omega\left(
1+\alpha^{2}\right)  \right]  \, .
\label{Condition2}%
\end{equation}
Constraint~(\ref{Condition2}) reduces to~(\ref{Condition}) for $a=0$.

The SSPM solution of the modified LLG equation~(\ref{8.33}) can be calculated
using a perturbative method based on Melkinov integrals \cite{8-59}. The
theoretical analysis can be experimentally verified by measurements of voltage
oscillations related to the dynamics of the magnetization, as previously
discussed. We believe that the parameters of the SSPM solution for the
 field $\mathbf{B}$, obtained in this experimental setup, can be compared with
the constraint~(\ref{Condition2}). This should be a nontrivial test
of the formalism developed in the present work.

\section{Final remarks}
\label{final-remarks}

In this work, a classical-quantum correspondence for a pseudo-Hermitian system
with finite energy levels is proposed and analyzed. A dictionary connects
particles subjected to real and complex fields ($\mathbf{B}$ and $\mathbf{F}%
$), related by a canonical transformation. The quantization map ensures
Hermiticity of operators, whose symbols are real functions in the respective
pseudoclassical phase space. The commutativity of the quantization map relates
canonical transformations between symbols to unitary transformations between
the corresponding operators. In particular, the Hamiltonians associated with
$\mathbf{B}$ and $\mathbf{F}$ are real under their classic involutions, and
the corresponding operators are symmetric (or Hermitian) with respect to the
inner products of the Hilbert spaces whereupon they act. An important point in
our development is the notion that there is no fundamental distinction
between Hermitian and pseudo-Hermitian (Hamiltonian) operators, or even
between ordinary quantum mechanics and pseudo-Hermitian quantum mechanics for
that matter, as long as the relations summarized by Fig.~\ref{diagram} are
satisfied. That is, as long as the possibility of a metric redefinition which
reestablishes Hermiticity in quantum theory can be seen as a consequence of a
proper choice of coordinates in the pseudoclassical theory. The only
nontrivial physical statement is that some non-Hermitian operators can become
pseudo-Hermitian under certain regimes.

Furthermore, we show that there is a unique isometry between the Hilbert
spaces $(\mathbb{C}^{n},\langle\cdot\,,\cdot\rangle)$ and 
$(\mathbb{C}^{n},\langle\cdot\,,\cdot\rangle_{\eta})$ that preserves the representation of
the Clifford algebra chosen in both settings (real and complex), implying a
unique metric. A systematic way of constructing this metric is provided. In
addition, we apply the classical correspondence to the two-level quantum
system coupled to a complex field. For non-Hermitian Hamiltonians, this
correspondence describes damping and does not recover the classical equations
of motion. When the Hamiltonian is pseudo-Hermitian, this correspondence does
not imply damping and the classical equations of motion are recovered.
Indeed, a common claim in the literature is that a complex Hamiltonian for a spin-\nicefrac{1}{2} particle always leads to spin precession with damping. We have shown that this is not necessarylly true. Hamiltonians associated with complex fields, which are non-Hermitian with the usual inner product, do not necessarily generate damping.

As a concrete development, we propose a damped version of the Rabi setup, considering
a complex field associated with a non-Hermitian Hamiltonian. We identify the
parameter that controls the intensity of the imaginary part as the Gilbert
damping parameter of the Landau-Lifshitz-Gilbert equation. In this setup, we
find a specific configuration of the parameters where the damping is
completely suppressed. In this case, the classical correspondence describes
again a precession movement for the spinning particle. We interpret this
arrangement as the configuration where the applied field completely
compensates the damping effect. It may be identified with the so-called
steady-state precession \cite{EvgTsZ2016}, where an external field cancels the
spin damping, generating a constant angle precession. The steady-state regime
could be observed with measurements involving ferromagnetic resonance methods
\cite{PetBaT2007}. We believe that the presented developments could be
verified in laboratory tests. For instance, using an experimental setup based on spin valves, the modified LLG constraint~(\ref{Condition2}) can be checked. This would be a nontrivial test of the theoretical formalism introduced.

The classical-quantum correspondence for two-level pseudo-Hermitian systems
may have practical applications. Precise manipulation of the spin has
several technological consequences and a description of damping process is
essential in this manipulation. For example, in the emerging technologies of
spintronic devices. In nowadays applications, the dynamics of the
magnetization in the digital storage process is described by the LLG equation
and any deviation from this description should have practical implications.
The possibility of suppressing the damping behavior could lead to a faster and
more energy-efficient spin manipulation. Phenomena in the steady-state
precession regime also have consequences in processes involving magnetic
resonance \cite{Car1958}.

Recent developments of the pseudo-Hermitian setup suggest interesting
perspectives for the theoretical framework presented here. Effects involving
non-Hermiticity enhances the dynamics of the topological-phase transitions,
bringing up new effects considering the scenarios involving the usual
Hermitian framework \cite{Kawabata}. Topological properties of the theory can
be explored, by evaluating quantities such as the Berry phase. A
second-quantization approach of the semiclassical damped Rabi problem proposed
in the present work can be investigated following a treatment in the same
lines as the one presented in Ref.~\cite{Bal17}. 

Finally, the developed
formalism might be extendable to lattice systems \cite{jin}. In this case
topological phase transitions in the exceptional points could be investigated,
as done for optical lattices \cite{ohashi}, as well as Anderson localization
and mobility edges in non-Hermitian systems \cite{ZenYo20}. 
Another topological phenomenon that can be explored in this context is the non-Hermitian skin effect \cite{NobKoKM20}. In this case, the suppression of the damping,
described in this work, may have some relationship with the elimination of the
non-Hermitian skin effect described in Ref.~\cite{ZenYa20}.

\begin{acknowledgments}
K.~R. acknowledges the support of Coordena\c{c}\~{a}o de Aperfei\c{c}oamento de Pessoal de N\'{\i}vel Superior (CAPES), Finance Code 001, Brazil; and
the National Council for Scientific and Technological Development (CNPq), Brazil, with grant \#141264/2020-9.
R.~F. acknowledges the support of S\~{a}o Paulo Research Foundation (FAPESP), Brazil, with grant \#2016/03319-6.
C.~M. acknowledges the support of National Council for Scientific and Technological Development (CNPq), Brazil, with grant \#420878/2016-5.
\end{acknowledgments}

\appendix

\section{Pseudoclassical model}
\label{appendix-pseudoclassical}

In this appendix we give a brief presentation of a simple nonrelativistic
model for a spinning particle in the context of pseudoclassical mechanics.
Following \cite{Ber77}, one considers a phase-space formulation where
dynamical variables are functions on a Grassmann algebra, such that, upon
quantization, their Poisson brackets provide the correct commutation relations.

The Grassmann algebra $G_{n}(\xi)$ is an algebra over the complex field
$\mathbb{C}$ whose generators $\xi_{i},~i=1,...,n$ satisfy the relations
\begin{equation}
\xi_{i}\xi_{j}+\xi_{j}\xi_{i}=0\,.
\end{equation}
Functions $f(\xi)$ on $G_{n}(\xi)$ are polynomials in the generators $\xi_{i}%
$. Hence, one can define a derivative operator acting on monomials and, by
extension, on functions, as the right-derivatives
\begin{equation}
\frac{\partial}{\partial\xi_{i}}\xi_{i_{1}}\xi_{i_{2}}\cdots\xi_{i_{k}}%
=\sum_{j=1}^{k}(-1)^{k-j}\delta_{ii_{j}}\xi_{i_{1}}\xi_{i_{2}}\cdots\xi
_{i-1}\xi_{i+1}\cdots\xi_{i_{k}}\,. \label{right-derivatives}%
\end{equation}

For our purposes, it is enough to consider Grassmann algebras with three
generators, that is, $n=3$. Thus, a general function $f(\xi)$ on the Grassmann
algebra $G_{3}(\xi)$ is given by
\begin{equation}
f(\xi)=f_{0}+f_{i}\xi_{i}+f_{ij}\xi_{i}\xi_{j}+\frac{\mathrm{i}}{3!}%
k_{f}\varepsilon_{ijk}\xi_{i}\xi_{j}\xi_{k}\,, \label{general-f}%
\end{equation}
where $f_{0},f_{i},f_{ij},k_{f}\in\mathbb{C}$ and $f_{ij}=-f_{ji}$. Odd-parity
functions $f$ are sums of homogeneous terms with odd numbers of the Grassmann
generators, and we write $P_{f}=1$. Even-parity functions $f$ are those
containing even number of generators, and we write $P_{f}=0$.

A relevant nonrelativistic pseudoclassical model is given by the action
\begin{equation}
S=\int_{t_{i}}^{t_{f}}L\left(  \xi,\dot{\xi}\right)  \,dt\,,\,L=\frac
{\mathrm{i}}{2}\xi_{i}\dot{\xi}_{i}-H\left(  \xi\right)  \,, \label{Action}%
\end{equation}
where $H(\xi)$ is some even function of the $\{\xi_{i}\}$, $P_{H}=0$. One can
proceed as in usual mechanics, and define the conjugate momenta
\begin{equation}
\pi_{i}=\frac{\partial L}{\partial\dot{\xi}_{i}}=\frac{\mathrm{i}}{2}\xi
_{i}\,, \label{momenta}%
\end{equation}
with the derivatives always taken from the right, as defined in
Eq.~(\ref{right-derivatives}). As a result, one finds the canonical
Hamiltonian function
\begin{equation}
H_{c}(\xi,\pi)=H(\xi)+(\pi_{i}-\frac{\mathrm{i}}{2}\xi_{i})\dot{\xi}_{i}\,.
\end{equation}

There is a natural Poisson bracket in the coordinates $(\xi,\pi)$. Let $f$ and
$g$ be functions of the Grassmann variables of definite parity. Then, the
Poisson bracket between them is defined as%
\begin{equation}
\left\{  f,g\right\}  =\frac{\partial f}{\partial\xi_{i}}\frac{\partial
g}{\partial\pi_{i}}-\left(  -1\right)  ^{P_{f}P_{g}}\frac{\partial g}%
{\partial\xi_{i}}\frac{\partial f}{\partial\pi_{i}}\,,
\end{equation}
where derivatives are taken from the right as usual. Thus, the Poisson
brackets between the canonical pairs are
\begin{equation}
\{\xi_{i},\pi_{j}\}=\{\pi_{j},\xi_{i}\}=\delta_{ij}\,.
\end{equation}
It should be noticed that the equations~(\ref{momenta}) are constraints, which
we denote as
\begin{equation}
\phi_{i}=\pi_{i}-\frac{\mathrm{i}}{2}\xi_{i}\, . \label{constraints}%
\end{equation}
Their conservation in time fixes the velocities $\{\dot{\xi}_{i}\}$:%
\begin{equation}
\{\phi_{i},H_{c}\}=0\implies\dot{\xi}_{i}=\mathrm{i}\frac{\partial H}%
{\partial\xi_{i}}\,. \label{derivadaxi}%
\end{equation}

Therefore, according to Dirac's terminology for constrained systems, the
model~(\ref{Action}) is a second-class theory, that is, there are no
first-class constraints and the dynamics is completely determined on the
constraint surface $\phi=0$. Following Dirac's quantization scheme for
second-class theories \cite{gitman1991}, we first define the Dirac brackets
over the set of second-class constraints $\phi$ as
\begin{equation}
\{f,g\}_{D(\phi)}=\{f,g\}-\{f,\phi_{i}\}C^{-1}_{ij}\{\phi_{j},g\}\,,
\end{equation}
where $C^{-1}_{ij}$ denotes the inverse matrix to $C_{ij}=\{\phi_{i},\phi
_{j}\}=-\mathrm{i}\delta_{ij}$, and again $f$ and $g$ are parity-definite
functions of the Grassmann variables. Thus the nonvanishing Dirac brackets
between canonical variables are
\begin{equation}
\left\{  \xi_{i},\xi_{j}\right\}  _{D(\phi)}=-\mathrm{i}\delta_{ij}\,,\left\{
\pi_{i},\pi_{j}\right\}  _{D(\phi)}=\frac{\mathrm{i}}{4}\delta_{ij}\,,\left\{
\xi_{i},\pi_{j}\right\}  _{D(\phi)}=\frac{1}{2}\delta_{ij}\,. \label{DB}%
\end{equation}
One can use the constraints $\phi$ to eliminate the momenta from the
description, so that one is only left with coordinates $\xi_{i}$.

\section{Quantization}
\label{appendix-quantization}

Exceptionally in this appendix we restore $\hbar$. Let us define a quantization map
$Q:G_{3}(\xi)\rightarrow L(\mathcal{H})$, where $G_{3}(\xi)$ is the Grassmann
algebra with generators $\{\xi_{i}\}_{i=1}^{3}$ and $L(\mathcal{H})$ is the
set of bounded linear operators on the Hilbert space $\mathcal{H}%
=(\mathbb{C}^{2},\langle\cdot,\cdot\rangle)$. It suffices to define the map on
monomials, following the antisymmetrization rule
\begin{equation}
Q\left(  \xi_{_{1}}\xi_{_{2}}\cdots\xi_{_{n}}\right)  =\frac{1}{n!}%
\sum_{\text{perm}}\left(  -1\right)  ^{\sigma\left(  \text{perm}\right)
}Q\left(  \xi_{i_{1}}\right)  Q\left(  \xi_{i_{2}}\right)  \cdots Q\left(
\xi_{i_{n}}\right)  \,,\label{symmetrization}%
\end{equation}
and extend it linearly to all functions. Furthermore, the quantization map $Q$
is required to map the unit to the identity in $\mathcal{H}$, $Q(1)=\mathbb{I}%
$. In our case, the above requirements imply the following for general
classical functions:
\begin{equation}
Q(f)=f_{0}\mathbb{I}+f_{i}Q(\xi_{i})+f_{ij}Q(\xi_{i})Q(\xi_{j})+\mathrm{i}%
\frac{1}{3!}k_{f}\varepsilon_{ijk}Q(\xi_{i})Q(\xi_{j})Q(\xi_{k})\,.
\end{equation}

It should be noticed that the quantization map $Q$ satisfies
\begin{equation}
f=f^{\ast}\Longrightarrow\langle x,Q\left(  f\right)  y\rangle=\langle
Q\left(  f\right)  x,y\rangle\,.
\end{equation}
That is, for real functions $f$, $Q(f)$ is symmetric, $Q^{\dagger}(f)=Q(f)$.
The map $Q$ is also required to satisfy the correspondence principle
\begin{equation}
\{f,h\}_{D(\phi)}=\lim_{\hbar\rightarrow0}\frac{1}{\mathrm{i}\hbar
}[Q(f),Q(h)]\,,
\end{equation}
where $[\cdot\,,\cdot]$ is a $Z_{2}$-graded commutator:
\begin{equation}
\lbrack Q(f),Q(h)]=Q(f)Q(h)-(-1)^{P_{f}P_{h}}Q(h)Q(f)\,,
\end{equation}
for all homogeneous functions $f$ and $h$. Thus, one has for the dynamical
variables $\{\xi_{i}\}_{i=1}^{3}$ the basic anticommutation relations
\begin{equation}
\lbrack Q(\xi_{i}),Q(\xi_{j})]=\hbar\delta_{ij}\,.\label{AC}%
\end{equation}
The above development is standard procedure on the program of quantization of
pseudoclassical models. Details can be found, for example, in reference
textbooks \cite{gitman1991,BerSh91}.

\end{document}